# Ferroelectricity driven magnetism at domain walls in LaAlO$_3$/PbTiO$_3$ superlattices

P. X. Zhou[1,2], S. Dong[3,a)], H. M. Liu[1], C. Y. Ma[1], Z. B. Yan[1], C. G. Zhong[2] and J. -M. Liu[1 b)]

[1]*Laboratory of Solid State Microstructures and Collaborative Innovation Center of Advanced Microstructures, Nanjing University, Nanjing 210093, China*

[2] *School of Science, Nantong University, Nantong 226007, China*

[3]*Department of Physics, Southeast University, Nanjing 211189, China*

[**Abstract**] Charge dipole moment and spin moment rarely coexist in single-phase bulk materials except in some multiferroics. Despite the progress in the past decade, for most multiferroics their magnetoelectric performance remains poor due to the intrinsic exclusion between charge dipole and spin moment. As an alternative approach, the oxide heterostructures may evade the intrinsic limits in bulk materials and provide more attractive potential to realize the magnetoelectric functions. Here we perform a first-principles study on LaAlO$_3$/PbTiO$_3$ superlattices. Although neither of the components is magnetic, magnetic moments emerge at the ferroelectric domain walls of PbTiO$_3$ in these superlattices. Such a twist between ferroelectric domain and local magnetic moment, not only manifests an interesting type of multiferroicity, but also is possible useful to pursuit the electrical-control of magnetism in nanoscale heterostructures.

**Keywords:** ferroelectric domain wall, *n-n* type superlattice, magnetoelectric coupling.

---

a) Corresponding author, E-mail: sdong@seu.edu.cn

b) Corresponding author, E-mail: liujm@nju.edu.cn

**Introduction**

With the developments of modern thin film technologies, it becomes possible to fabricate and characterize oxide superlattices (SLs) and heterostructures (HSs) down to the atomic level[1-3]. Oxide SLs and HSs can display many curious properties different from the parent bulk materials. For example, those SLs constructed by charge-polar layers and charge-neutral layers, have attracted a lot of research interests for the two-dimensional electron gas (2DEG) due to the so-called polar catastrophe[4-6]. Recently, theoretical simulations predicted that if ferroelectrics are involved in the SLs with charge polar discontinuity, the ferroelectricity can modulate the interfacial 2DEG, namely the 2DEG can be switched from one interface to the opposite one by flipping the ferroelectric (FE) polarization[7-11].

It is the common sense that ferroelectricity (or antiferroelectricity) and ferromagnetism (or antiferromagnetism) are difficult to coexist in one system because in general FE dipoles favor empty $d$ orbitals while magnetic moments need partially-occupied $d$ orbitals. Therefore, despite the great progress in the past decade, the magnetoelectric performance of most multiferroic bulks remains poor. An alternative route to pursuit magnetoelectricity is to utilize oxide HSs, e.g. a strong FE material plus a strong ferromagnetic (FM) material. The direct advantage of such magnetoelectric HSs is the strong FE and FM signals, however, the magnetoelectric coupling between the FE and FM layers is indirect, which is usually via the strain effect or field effect. Very recently, Rogdakis *et al.* obtained the FE polarization in the NdMnO$_3$/SrMnO$_3$/LaMnO$_3$ SL[12], in which the FE polarization was related to the antiferromagnetic (AFM) order of Mn. In this sense, the NdMnO$_3$/SrMnO$_3$/LaMnO$_3$ SL is somewhat similar to the so-called type-II multiferroics, in which the FE polarizations are directly driven by magnetic orders and thus the strong magnetoelectric coupling is intrinsically guaranteed[13,14].

In this work, another kind of oxide SLs consisting of LaAlO$_3$ (LAO) and PbTiO$_3$ (PTO) is investigated to pursuit the inverse property: the polarization driven magnetism. Both LAO and PTO are well studied perovskites. The choice of PTO and LAO has several reasons. First, LAO is a frequently used substrate with polar terminations [(LaO)$^+$ and (AlO$_2$)$^-$] along the [001] direction, against the charge-neutral layers of PTO [(PbO)$^0$ and (TiO$_2$)$^0$]. Thus the charge polar discontinuity exists in these SLs which may induce the 2DEG, similar to the situation in extensively studied LaAlO$_3$/SrTiO$_3$ (STO)[4-6,15-18]. Second, PTO is a typical FE material including FE-active Ti$^{4+}$ and Pb$^{2+}$. Thus PTO has a very strong FE polarization along the *c*-axis ([001] direction), which will be further enhanced by the in-plane compressive strain from LAO. Such a robust ferroelectricity is advantageous to compete with the polar discontinuity and modulate the 2DEG in SLs, while there's no ferroelectricity involved for 2DEG in LAO/STO cases. Third, LAO is a band insulator with a large band gap (5.6 eV)[4], which is larger than the band gap of PTO (3.6 eV)[19]. Thus, the band alignment between PTO and LAO will restrict the 2DEG in the PTO side. Fourth, the nominal valence of Ti in perovskites can vary from +4 as in PTO to +3 as in YTiO$_3$ (the latter is magnetic-active), which allows the possibility to "create" magnetic moments in these SLs. Last, despite the non-negligible lattice mismatch, previous experiments have fabricated epitaxial PTO films on the LAO substrates[20] and there are also theoretical studies on LAO/PTO SLs[21], making the present design be practical.

Regarding the interfaces, there are three types in these (LAO)$_m$/(PTO)$_m$ SLs. The first one includes both the *n*-type (LaO)$^+$/(TiO$_2$)$^0$ and *p*-type (AlO$_2$)$^-$/(PbO)$^0$ interfaces, as shown in Fig.1 (a), which keep the stoichiometry of LAO and PTO but break the symmetry between the two interfaces. The second one is with double *n*-type (LaO)$^+$/(TiO$_2$)$^0$ interfaces, which can be denoted as *n*-*n* type interfaces. The third one is with double *p*-type (AlO$_2$)$^-$/(PbO)$^0$ interfaces. It is noted that the latter two cases break the stoichiometry of original LAO and PTO, which will bring more carriers (electrons or holes) to the systems. The non-stoichiometry brought by the double *p*-type interfaces may result in oxygen vacancies[5] although it remains under debate, which makes the modeling complex and also is not the intrinsic behavior of interest in this work. Therefore, in the present study, we will only focus on the *n*-*p* type and *n*-*n* type SLs, and leave the double *p*-type SLs to further investigations. Benefited from the advancement of experimental techniques, it becomes possible to fabricate these superlattices with deserved interfaces (*n*-*n*, *n*-*p*, or *p*-*p*) precisely, and the ion vacancies (e.g. oxygen vancanies) can be reduced to a minimal level. In the following studies, each *n*-*p* type SL and a double *n*-type SL are indexed by an integer *m* and a half integral *m* respectively.

The magnetoelectric properties of the (LAO)$_m$/(PTO)$_m$ SL with either integer or half-integer *m* have been investigated. Our results show that the double *n*-type *m*=4.5 SL is the most interesting one for the magnetoelectric coupling. For the *n-p* interface SL at *m*=4, it is predicted that the PTO layer become a polarized single-domain, no matter what the initial state of the PTO layer is. In this case, the SLs have no magnetism. However, for the *n-n* interface SL, several emergent phenomena are predicted. First, the internal electric field arisen from the (LaO) polar layer pointing to the doped electron in $TiO_2$ layers makes PTO favor a head-to-head FE domain if *m* is no more than 3.5. For *m*>3.5, the domain structure of the PTO layers depends on the initial state. More importantly, local magnetism is generated and the maximal moment appears at the domain wall. This feature allows a possible control of the position of magnetism by external electric field via the ferroelectric polarization switching, which may be useful to design a domain-wall memory. Here, an additional magnetoelectric mode accounting for the dot product of the polarization spatial gradient and squared magnetic moment is predicted.

**Results**

**The *m*=4 SL with the *n-p* type interfaces**

First, we start from the simplest case: the *m*=4 SL which keeps the stoichiometry of LAO and PTO layers, as sketched in Fig.1(a). In our calculations, three types of initial conditions for PTO layers: one paraelectric (PE) state, two FE states with a positive/negative polarization, have been adopted for structure relaxation. However, the relaxed SL structures evolved from all the three initial conditions become identical, which are all unidirectional polarized with the polarization pointing from the *n*-type interface to the *p*-type one, as shown in Fig.1(a), in agreement with previous studies on a thinner LAO/PTO SL[21]. The local polarization can be characterized by the cation-anion displacement (denoted as *D*) of each layer, as shown in Fig.1(b), which suggests a single polarized domain of the whole PTO. This polarized domain is very robust despite the different initial conditions, which can be understood as the result of an internal electric field (*E*) built by the asymmetric *n*- and *p*-type interfaces, as proposed in previous studies[22] and illustrated in Fig.1(c). This internal electric field, which can be roughly estimated as 20 MV/cm by modeling the SL as a simple capacitor, far beyond the typical coercive field of PTO, and thus can polarize the initial PE state to the final polarized single domain and even reverse the domain if the initial polarization is in the opposite direction.

The internal electric field within the PTO layers can be also evidenced from the projected density of states (PDOS) of each layer, as shown in Fig.1(d). According to the density of states (DOS), the SL remains insulating, but the bottom of conducting band and top of valence band shift between layers. The amplitude of band shift between two interfaces is about 1.5 eV. It should be noted that the shift direction is in opposite to the electric field $E$. This is due to the depolarization field of PTO itself, which is even larger than the one from polar terminations. According to our DFT calculation, the polarization of PTO bulk on the LAO substrate is 99.75 $\mu C/cm^2$, which is equivalent to ~0.9 electron per unit cell surface and larger than the 0.5 electron per $n$-type termination. Furthermore, this fact can be easily checked by fixing the initial PE structure of PTO (thus no polarization nor depolarization field) in the calculation, then the PDOS shift (not shown here) is in opposite to that in Fig.1(d). In other words, the electric field $E$ from the polar interfaces provides a preferred orientation (from left to right as in Fig.1) for the spontaneous polarization of PTO layers. Then the depolarization field from the final PTO structure is in opposite to the polarization, i.e. from right to left. Since the spontaneous polarization of PTO is very large, the depolarization field overcomes the $E$ from polar interfaces, making the potential energy of electrons in PTO layers lower in the $p$-type termination side than the $n$-type side.

All above calculations were done without magnetism. Subsequently, the magnetism is switched on. However, the initial magnetic moments ($M$) in this SL quench to zero after the electronic self-consistent convergence. In short, this $m$=4 SL only owns a single polarized domain and is non-magnetic. This non-magnetic result agrees with Fig.1(d): there is no occupation of the Ti's 3$d$ band.

**The $m$=4.5 SL with the $n$-$n$ type interfaces**

Since the $m$=4 SL addressed above is somewhat trivial, we turn to the $m$=4.5 SL with double $n$-type interfaces, which is more interesting. In this SL, the non-stoichiometry is created by the extra LaO layer: namely there are five $TiO_2$/LaO sheets but only four PbO/$AlO_2$ sheets, as shown in Fig.2(a). The symmetry is maintained by such $n$-$n$ interfaces.

Using the same procedures, both the PE and FE initial conditions are imposed to the PTO

layers. Different from the case of *m*=4 SL, here the relaxed structures are no longer unique but depends on the initial conditions. If the initial structure of PTO is in the symmetric PE phase, the relaxed structure displays centrosymmetric domains with a head-to-head domain wall at the central PTO layers, as indicated by the blue symbols in Fig.2(b). The local dipole, namely the layer-dependent anion-cation displacement, decreases from the two interfaces to the central $TiO_2$ layer. The origin of such a domain structure can be straightforwardly understood as the effect of polar interfaces. In the *n-n* type SL, there are two LaO terminations, each of which donates 1/2 electron to the PTO layers. In other words, each LaO termination has +0.5 unit charge/unit cell. Thus an inner electric field is established, going from the LaO termination to the compensating doped electron in $TiO_2$ layers, as sketched in Fig 2(d). It is the symmetric inner field which polarizes the PTO and thus results in the head-to-head domain wall at the center of PTO.

In contrast, when the initial structure of PTO is in the FE phase, the relaxed structure displays asymmetric FE domains, with a dominant one pointing along the initial direction and a small opposite one located at one interface, as indicated by the red symbol in Fig.2(b). The domain wall remains head-to-head but is no longer at the center of PTO. The intensity of layer-dependent dipole also decreases from the interface to the domain wall. This domain structure can be understood as the competition between the inner electric field (established between polar LaO terminations and doped electron) and the spontaneous ferroelectricity of PTO. As sketched in Fig.2(e), PTO itself owns a robust FE polarization, e.g. from left to right. Then the left LaO termination further enhances this FE polarization. However, the right LaO termination prefers to polarize the PTO layers to the opposite direction. This frustration between the inner electric field and FE polarization generates such an asymmetric FE domain structure.

The energy of the asymmetric domain (with the FE initialization) is lower than the symmetric one (with the PE initialization) for about 11-28 meV/per Ti. Even though, the symmetric domain remains meta-stable during the structure relaxation, which can play as the transition state connecting the positive/negative asymmetric FE domains.

Then magnetism is considered due to possible magnetic moment of $Ti^{3+}$ in such a *m*=4.5 SL. Both the FM and A-type AFM orders are taken as the initial spin structures together with the PE and FE conditions to do the structure relaxations, while the outputs after the electronic self-consistent convergence favor the FM profile. Interestingly, the involvement of magnetism does not change the displacements of ions too much, which are almost identical to those in the non-magnetic ones, as compared (solid- *vs* open-symbols) in Fig.2(b). This implies a negligible magnetoelastic coupling in the PTO layers, in agreement with previous DFT study of magnetic titanates[11,23].

The magnetic moment of each ion, calculated using the Wigner-Seitz sphere as specified by VASP, is shown in Fig.2(c). Meanwhile, the total moment of superlattice is calculated directly as the occupy difference between the spin-up electrons and spin-down electrons in the whole system. Given both the PE and FE initial conditions, magnetism emerges in the PTO layers and the total magnetic moment reaches 0.208 $\mu_B$ and 0.662 $\mu_B$ (without Hubbard $U$) respectively. It is reasonable that the total magnetic moment is between 0 $\mu_B$ and 1 $\mu_B$ because there is only one doped electron in the PTO part. The details of local magnetic moments with and without Hubbard $U$ are summarized in Table I. The Hubbard $U$ will not change the results qualitatively, although it enhances the local magnetic moments as expected.

As mentioned above, the doped electron can be trapped around the head-to-head domain wall in PTO layers, thus the largest local magnetic moments in both the PE and FE initial conditions always appear at the domain walls, despite the different domain wall positions. Although the charge dipole and magnetism are mutually exclusive, the local magnetism can survive at the head-to-head domain walls as an edge state effect of ferroelectrics, since the local FE displacement vanishes at such domain walls. This domain wall carried (local) magnetic moment is switchable electrically, rending a magnetoelectric functionality. Namely, if the domain wall is switched between the PE and FE-type ones or between the positive and negative FE-type ones by external electric fields, the peak of magnetic moments will move among the left, middle, and right of PTO layers.

It should be noted that although the total magnetic moment is a constant before and after such a switch, the movement of magnetism carried by the FE domain can still be considered as a type of microscopic magnetoelectric coupling. In a famous review on multiferroics by Khomskii[13], the local ferroelectric dipole originated from the Néel-type ferromagnetic domain wall was highlighted, which was just a counterpart of our case. In addition, in recent years, the so-called racetrack memory or domain-wall memory attracted many research attentions, in which the ferromagnetic domains are moved by electric current. Furthermore, during the polarization flipping, the transition state, e.g. the so-called PE state in our calculation, will show different magnitude of magnetization. Thus, a magnetoelectric dynamic progress is naturally expectable, e.g. electronmagnon driven by the polarization flipping.

**Thickness effect of SLs with the *n-n* type interfaces**

It is well known that FE domain structures seriously depend on the size of samples. In the SL geometry, the thickness effect of PTO layers deserves a systematic check. According to the above calculations, the induced magnetism is always associated with the FE domain wall in the SL with the *n-n* type interfaces. Thus, here we only focus on those SLs with the *n-n* type interfaces: *m*=1.5, 2.5, and 3.5.

In all these thinner SLs, the initial PE and FE conditions give the same relaxed structure, with the head-to-head domain at the center layer, as shown in Fig.3(a). This domain structure implies that the spontaneous ferroelectricity is overwhelmed by the internal built electric field from the polar interfaces and the depolarization field of PTO itself. The former one is nearly a constant for all thickness SLs but the later one is more prominent in thinner PTO layers. The induced magnetic moments of Ti ions are similar to the corresponding situation of the *m*=4.5 case, as listed in Table I.

The total magnetic moments of the *n-n* type SLs are displayed in Fig.3(d) as a function of the periodic thickness *m*. It is obvious that the total magnetic moments are significantly enhanced with the decreasing thickness of those SLs when the pure GGA method is used, which is reasonable since the average valences of Ti ions in thinner PTO cases are closer to the magnetic-active $Ti^{3+}$. While using the GGA+*U* method, the total magnetic moment is close to the saturated value (1.0 $\mu_B$ due to the non-stoichiometry), and thus is almost independent of the thickness. These results agree with the scenario of FE domain wall carried magnetism. However, in these SLs, the domain walls and their affiliated magnetism are not electrical switchable.

Besides above thinner *m* cases, a thicker case is also tested. According to our above results, the LAO layers play three roles. First, the LaO terminations dope one extra electron to the PTO layers in the *n-n* type SLs. Second, the polar terminations build an inner electric field. Third, the highly-insulating LAO barrier confirms the extra carrier within the PTO layers. However, the cation $Al^{3+}$ itself is neither ferroelectric-active nor magnetic-active, and thus the thickness of LAO layers, if not too thin, will not qualitatively affect the ferroelectricity and magnetism in the PTO layers. Thus, to save the computational CPU time, here we calculate $(LAO)_{3.5}/(PTO)_{6.5}$, in which the thickness of LAO does not equal to that of PTO, which will not alter the physical conclusion. The relaxed crystal structures (Fig.3(b)) are similar to those of the *m*=4.5 SL, giving two types of domain walls for the PE and FE initial conditions. The layer-dependent magnetic moments of Ti ions in this SL are shown in Fig.3(c) and also listed in Table I. In agreement with previous studied cases, the maximum magnetic moments appear at the domain walls, although the total magnetic moment is seriously suppressed in the PE initial condition without Hubbard *U*.

By comparing the data of these SLs, it is found that *m*=4.5 is the critical and optimal thickness for the PTO layers in the SLs to maintain switchable FE polarization and domains, which is useful for designing the magnetoelectric devices. Although the thinner periodic SLs may own larger magnetic moments, the symmetric domain structure is too robust to be switched directly. For thicker periodic SLs, although the switch function is available, the magnetization per volume is weak, since the total magnetization is proportional to the number of superlattice period per unit volume. Given fixed thickness of total PTO layers, the *m*=4.5 one is the optimal choice to keep switchable FE polarization and a maximum magnetization.

**Further verification**

It is recognized that SLs with polar interfaces are quite complicated and GGA/GGA+*U* method may give "pathological" band alignment at the interface[24-26]. Thus it is essential to pay special attention to the results as well as the calculation technique.

First, from the physical consideration, here the charge transfer found in the DFT calculation is reasonable, in agreement with the expectation as stated in the introduction section. In such SLs, only the Ti ion can change its valence (from +4 to +3) easily, while the valences of other elements are almost constants. Thus, the extra electron from the *n-n* polar interfaces is naturally expected to be confined in the PTO layer. This scenario also agrees with the LAO/STO interfaces, where the extra electron stays in the STO side[5].

Second, the hybrid functional calculations with Heyd-Scuseria-Ernzerhof (HSE) exchange correlation[27-29] is performed to verify the GGA and GGA+$U$ results. Since the HSE calculation is extremely CPU-time consuming, only the shortest periodic SL ($m$=1.5) and parent materials are tested, while others are beyond our computational capabilities. According to the band gaps shown in supplementary, the HSE indeed gives a better description of PTO and LAO. Then, the crystal structure of SL is relaxed and the magnetic property is calculated. Despite the initial state (PE or FE), the relaxed structure remains the centrosymmetric head-to-head domain, perfectly coinciding with the GGA/GGA+$U$ one. The magnetic moments are 0.472$\mu_B$/Ti, also in consistent with the results of GGA+$U$.

Third, generally, the head-to-head FE domain is not usual since it is energetically costly. However, previous literature reported this type of domain in PTO[30-32]. Especially, it is highly possible in the present SLs since there's a sheet of negative charge between the domains which compensates the electric fields. In addition, when approaching the domain wall, the local polarization is progressively reduced (to zero), which further reduces the electrostatic energy. In fact, this head-to-head FE domain is stabilized by the inner electric field. Without these polar terminations, the compensation of polarization may break the system into lateral domains[33-35], e.g., Aguado-Puente and Junquera showed SrRuO$_3$/BaTiO$_3$/SrRuO$_3$ capacitors could form closure domains. To verify the robust of the head-to-head structure, more calculations with expanded in-plane unit-cell ($\sqrt{2}\times\sqrt{2}$, 1×2, 1×4) have been done. Such expanded cells have been tested for both the $m$=2.5 and $m$=4.5 ($\sqrt{2}\times\sqrt{2}$ only) SLs. Starting from the proper 180$^0$ domains (like $\overleftrightarrow{\phantom{x}}$ )[33], both SLs are finally relaxed to the centrosymmetric head-to-head domain despite the technique details (GGA or GGA+$U$, and the length of $c$-axis), identical to the above PE one in the 1×1 in-plane unit cell. All these results suggest that the head-to-head domain is robust since the internal field induced by the polar interface is so strong that make the charge compensation dominant rather than the compensation of breaking up the system into lateral domains.

**Discussion**

The FE domain wall carried magnetism revealed above is not only potentially useful in applications, but also stands for an interesting type of magnetoelectric coupling. The energy of a system can be expressed as a function of FE polarization $P$ and magnetic moment $M$. From the symmetry consideration, the first-order coupling between $P$ and $M$, namely $P \cdot M$, is forbidden, because $P$ violates the space-inversion symmetry but $M$ violates the time-reversal symmetry, and the system energy should not change under the individual operation of time reversal or space inversion. The second order coupling $P^2M^2$ is allowed to manifest the magnetoelectric coupling, which can be achieved via the proximity strain effect between magnetostrictive materials and ferroelectrics with ferroelastic behavior.

The great progress of multiferroics in the past decade has revealed more types of magnetoelectric couplings. A typical example is the improper FE polarizations induced by noncollinear magnetic orders, e.g. in TbMnO$_3$ or at the Néel-type ferromagnetic domain wall. In these spiral magnets, the magnetoelectric coupling term in the energy is in the form of $P \cdot [M(\nabla \cdot M)-(M \cdot \nabla)M]$[36], which can also maintain both time-reversal and space-inversion symmetries but the order is lower than above $P^2M^2$. Thus this type of coupling can be essentially stronger than the conventional one, especially in terms of magnetic control of polarization. In addition, the electromagnon, namely a spin wave excited by an alternating electric field, is another type of magnetoelectric coupling, which can be expressed as a term like $dP/dt \cdot (\nabla \cdot M)$ (or its variations) in the energy[37].

The electromagnetic coupling revealed in the current work belongs to an additional one. In analogous to the above expression, the derivative of $P$ (e.g. $\nabla \cdot P$), can be used to couple with any terms of even order of $M$, e.g. $M^2(dP_z/dz)$ (or its variations) in the SL geometry where $z$ is the position vector along the stacking direction, which can also satisfy the symmetry requirements of energy. The relationship between local $dP_{iz}/dz$ and $M_i^2$ in all the above studied SLs (with pure GGA) is summarized in Fig. 4. It is obvious that the induced $M_i^2$ has a coupling with $dP_{iz}/dz$: a large $M_i^2$ always accompanies a large $dP_{iz}/dz$. This type of coupling, also with a lower order than the conventional $P^2M^2$, gives an intrinsically strong magnetoelectric coupling, especially in terms of electrical control of magnetism.

In fact, the term $dP_z/dz$, which means a net charge at position $z$, classifies this type of local magnetoelectric coupling to be the carrier driven one. This concept was proposed by Rondinelli, Stengel, and Spaldin[38]. Recent experimental and theoretical investigations on the FE field effect in FE-manganites heterostructures can be also classified to this type electromagnetic coupling[10,39-43]. Indeed, it was found that the termination of FE polarizations (large $dP_z/dz$) at the FE-manganite interfaces controls the interfacial magnetism.

Despite the similarity of our current work to these previous studies, there are still some conceptual differences. In all these previous FE-manganite HSs, the order parameters **P** and **M** belongs to two different materials, coupled only via the interface. In the present SLs, both the local $\boldsymbol{P}_i$ and $\boldsymbol{M}_i$ reside in the single component PTO. Then the coupling between $M_i^2$ and $dP_{iz}/dz$ can be locally on-site and thus more direct and stronger. The key differences of magnetoelectric coupling between these SLs and other systems are listed in Table II.

**Conclusion**

In summery, we have investigated the crystal structures and electronic structures of $(PTO)_m/(LAO)_m$ SLs. In the SLs with double $n$-type interfaces, the head-to-head FE domains are observed, whose structures are determined by the thickness of PTO layers and the initial FE or PE states. Such FE domains are understood as a result of the competition between the polar discontinuity and spontaneous FE polarization. The most interesting finding is that ferromagnetically aligned magnetic moments emerge at these FE domain walls. Such FE domain wall carried magnetism evades the principle of mutual exclusion between ferroelectricity and magnetism, since the local FE displacement at the head-to-head FE domain wall vanishes. In contrast, the $m$=4 SL with asymmetric $n$- and $p$-type interfaces, owns a single domain (no domain wall) polarized by the interfaces, which is magnetism free. Our calculation predicts that $m$=4.5 is a critical and optimal thickness for switchable FE domains and thus magnetoelectric function.

**Methods**

To explore the crystal and electronic structures of the $(LAO)_m/(PTO)_m$ SLs, a density functional theory (DFT) calculation is performed based on the Perdew-Burke-Ernzerhoff (PBE) exchange-correlation form of generalized gradient approximation (GGA) implemented in the Vienna *ab-initio* simulation package (VASP)[44,,45].

It is well known that DFT calculations often underestimate band gaps in insulating systems with localized *d*- and *f*-orbitals. Here, the band gaps of FE PTO and LAO are about 2.0 eV and 3.84 eV in our pure GGA calculation, respectively. Thus, following previous literature[46,47], the GGA+*U* method in Liechtenstein approach[48] with $U_d$=5 eV, $J_d$=0.64 eV on Ti's *d*-orbital is also adopted to compare with the pure GGA ones. Then the band gap of PTO increases to 2.3 eV, which is closer to the previous ones[49]. Since La's 4*f* bands lie higher than the GGA prediction, $U_f$=11 eV, $J_f$=0.68 eV is imposed on La's *f*-orbital to avoid mixing with the conduction band[46,50-53]. In addition, to double-check the possible underestimation of band gaps using GGA or GGA+*U* method which may result in (what Junquera and Ghosez call) "pathological" band alignment in heterostructure[24-26], the shortest SL and parent materials will be verified using the hybrid functional calculations. The energy cutoff of plane-wave basis is 500 eV in all calculations. The self-consistent calculations are converged to $10^{-6}$ eV/u.c., and the crystal structures are relaxed until the Hellman-Feynman forces on ions are all below 0.01 eV/Å. The Monkhorst-Pack *k*-point mesh is 10×10×1 for the *m*=4 and longer periodic SLs, and is increased for other shorter periodic SLs correspondingly.

All SLs studied here are supposed to grow along the conventional [001] direction. First, the structure of cubic LAO is fully relaxed, which gives the lattice constant as 3.811 Å, very close to the experimental value 3.81 Å[54] and previous DFT one 3.79 Å[55]. Then the in-plane lattice constants of PTO are fixed to this calculated value of LAO lattice, and its out-of-plane lattice constant *c* and the inner atoms' positions are fully relaxed. Both the PE state and FE state of PTO are considered in these relaxations. Since the in-plane lattice constant of LAO substrate is smaller than that of PTO (3.904 Å[56], FE state), the lattice constant of PTO along the *c*-axis is significantly elongated when using LAO as the substrate. To construct the SLs, the interfacial space between the two materials is determined by changing the distance between PTO and LAO without altering the atomic positions and the lattice constants, and then searching for the minimal energy of the total system[7,51]. The atomic positions and the *c*-axis of all SLs are fully optimized once more to obtain the final SLs structures.

To verify the rationality of the GGA and GGA+*U* methods, the comparisons of three more methods (LDA, LDA+*U*, HSE) are supplied in the supplementary.

**TABLE I.** The local magnetic moments of each Ti ion and total moments (in unit of $\mu_B$) in the initial PE and FE conditions of *n-n* type SLs. The local moments are calculated within the Wigner-Seitz sphere as defined in the PAW potential of VASP. The index of Ti can be read from figures. The total moment includes the contributions from other ions. The data in the first and second rows of each SL are obtained from the pure GGA and GGA+*U* calculations, respectively. For those FE/PE switchable cases, the energy difference (in unit of meV/Ti) between the PE and FE conditions is listed.

| Initial structure | Local | | | | | | | Total | Energy |
|---|---|---|---|---|---|---|---|---|---|
| | Ti1 | Ti2 | Ti3 | Ti4 | Ti5 | Ti6 | Ti7 | | |
| *m*=1.5 PE/FE | 0.354 | 0.354 | | | | | | 0.680 | |
| +*U* | 0.473 | 0.473 | | | | | | 0.856 | |
| HSE | 0.472 | 0.472 | | | | | | 0.852 | |
| *m*=2.5 PE/FE | 0.115 | 0.477 | 0.115 | | | | | 0.640 | |
| +*U* | 0.161 | 0.676 | 0.161 | | | | | 0.864 | |
| *m*=3.5 PE/FE | 0.032 | 0.294 | 0.294 | 0.040 | | | | 0.584 | |
| +*U* | 0.115 | 0.396 | 0.394 | 0.116 | | | | 0.853 | |
| *m*=4.5 PE | -0.006 | 0.055 | 0.144 | 0.053 | -0.001 | | | 0.208 | |
| +*U* | 0.099 | 0.176 | 0.504 | 0.177 | 0.098 | | | 0.865 | |
| *m*=4.5 FE | 0 | 0.001 | 0.058 | 0.429 | 0.237 | | | 0.662 | -11 |
| +*U* | 0 | 0.003 | 0.165 | 0.681 | 0.111 | | | 0.862 | -28 |
| LAO$_{3.5}$/PTO$_{6.5}$ | | | | | | | | | |
| PE | 0 | 0.001 | 0.015 | 0.052 | 0.006 | 0.001 | 0 | 0.062 | |
| +*U* | 0 | 0.061 | 0.154 | 0.623 | 0.150 | 0.006 | 0 | 0.835 | |
| LAO$_{3.5}$/PTO$_{6.5}$ | | | | | | | | | |
| FE | 0 | 0 | 0 | 0 | 0 | 0.130 | 0.491 | 0.605 | -80 |
| +*U* | 0 | 0 | 0 | 0 | 0.053 | 0.264 | 0.667 | 0.869 | -48 |

**TABLE II.** Manifestation of magnetoelectric couplings. The magnetoelectric coupling in the LAO/PTO SLs is identical to neither bulk multiferroics nor FE-manganite HSs (or analogous FE-magnetic ones).

| System | Hosts of *P* & *M* | Coupling |
|---|---|---|
| multiferroics | homo-material | spin-orbit/spin-lattice |
| FE-manganite | hetero-material | carriers |
| LAO/PTO | homo-material (PTO) | carriers |

**Acknowledgments**

Work was supported by the National 973 Projects of China (Grants No. 2011CB922101), the Natural Science Foundation of China (Grants Nos. 11234005, 11274060, 51332006, 51322206), and the Priority Academic Program Development of Jiangsu Higher Education Institutions, China.

**Author contributions**

S.D. and J.M.L conceived the idea. P.X.Z. performed the simulation. S.D., P.X.Z., and J.M.L analyzed the data. S.D. and P.X.Z wrote the manuscript. H.M.L, C.Y.M., Z.B.Y, and C.G.Z. discussed the results and commented on the paper.

*Figures*

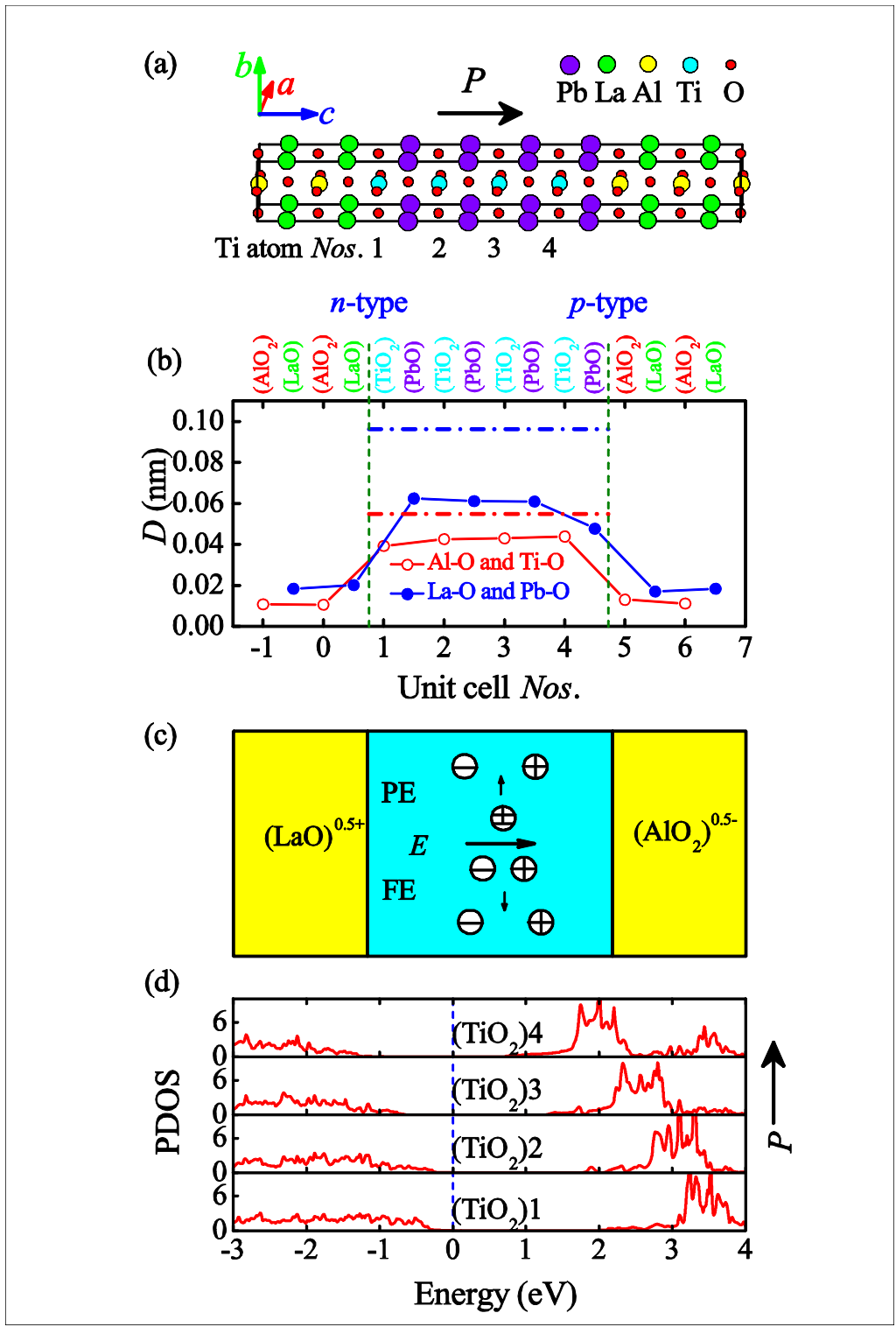

**Fig.1.** Results of the *m*=4 SL with *n-p* type interfaces. (a) Schematic crystal structure. (b) The local displacements between anions and cations. Both the PE and FE initial conditions give the same profiles. The vertical dash lines denote the two interfaces. For reference, the red and blue dot lines denote the Ti-O and Pb-O relative displacements of PTO bulk on LAO substrate, respectively. (c) Sketch of the mechanism. No matter the initial condition is PE or FE, the internal built electric field from the *n-p* interfaces can polarize the PTO layers. Circle with sign stands for the center of positive/negative charge. (d) The PDOS of each TiO$_2$ layer, which shows a clear band shift (1.5 eV) due to the polar *n-p* termination pair and depolarization field. The Fermi energy is positioned at zero, as indicated by the vertical dash line.

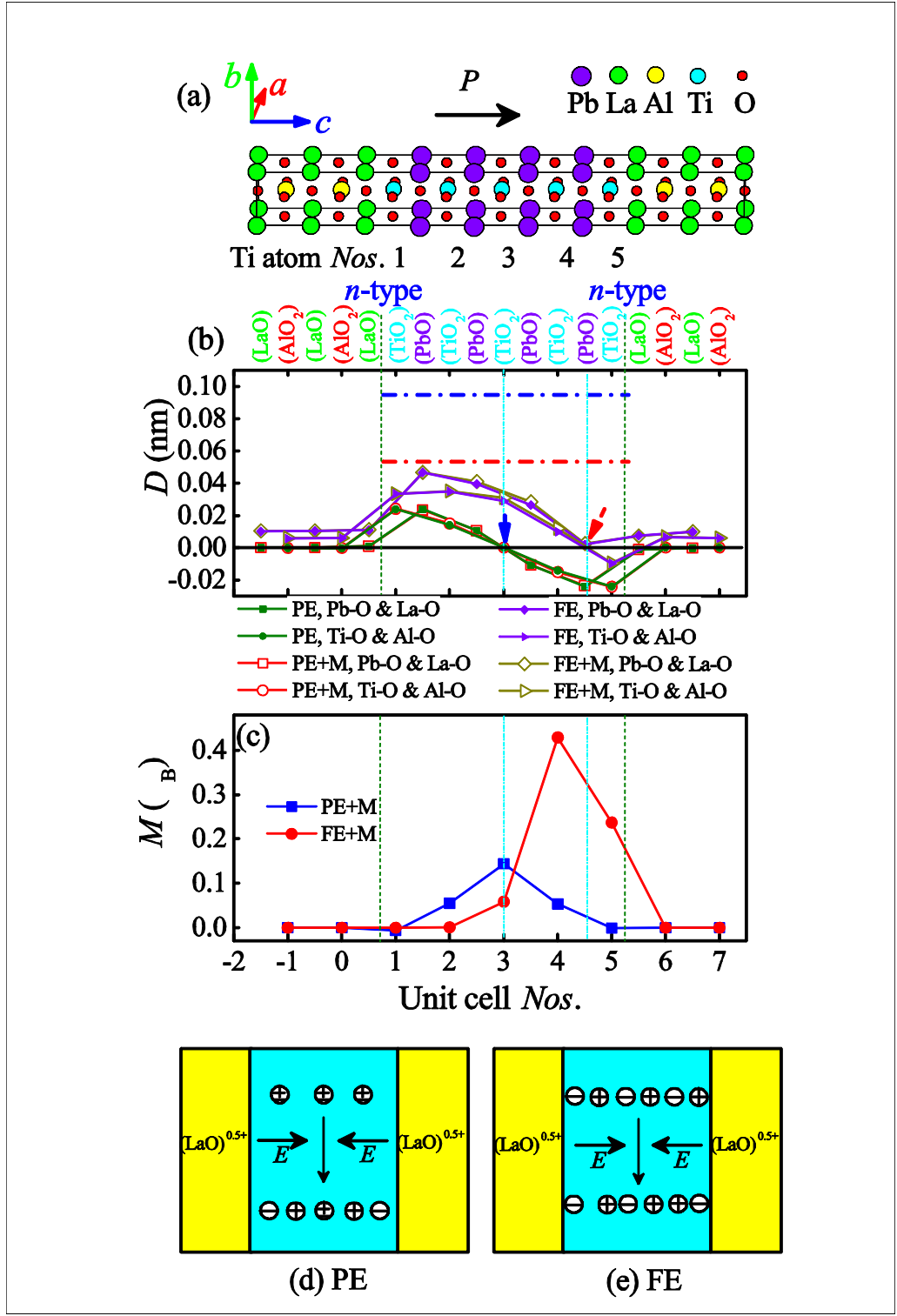

**Fig.2.** Results of the *m*=4.5 SL with *n-n* type interfaces. (a) Schematic crystal structure. The initial FE polarization points from left to right; (b) The local displacements between anions and cations. Both the PE and FE conditions with (+*M*)/without magnetism are shown by different symbols. The corresponding symbols with magnetism almost overlap with those without magnetism. The vertical olive dash lines and cyan dot lines denote the interfaces and domain walls, respectively. For reference, the red and blue dot lines denote the Ti-O and Pb-O relative displacements of PTO bulk on LAO substrate, respectively. (c) Local magnetic moments. Here only Ti atoms show finite magnetic moments. (d-e) Sketch of the mechanisms for the (d) PE and (e) FE conditions. The internal built electric fields from the two interfaces to the center of doped electron tune the FE domains.

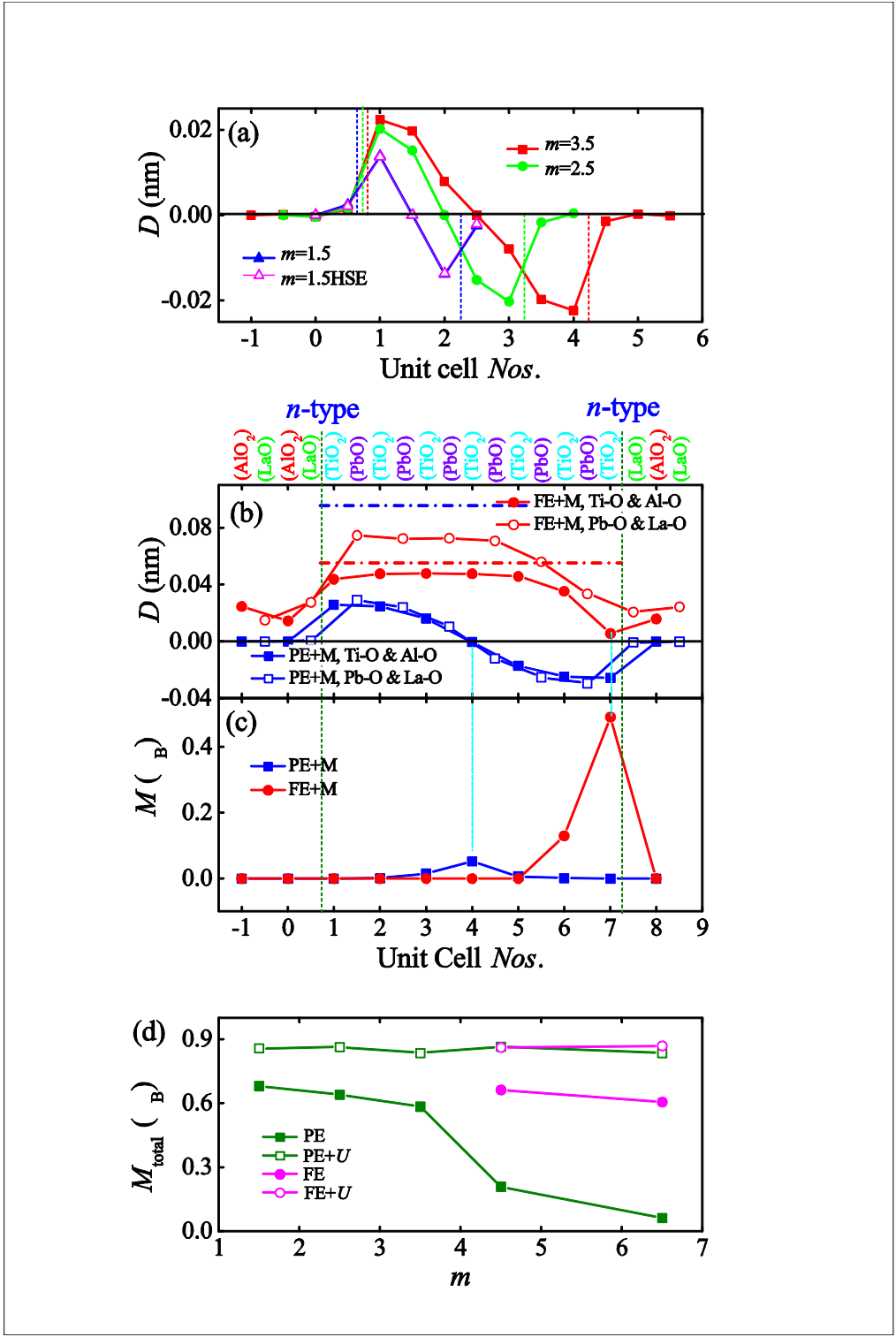

**Fig.3.** (a) The local displacements between anion and cation ions for *m*=1.5, 2.5 and 3.5 SLs with *n-n* type interfaces. The vertical lines denote the position of interfaces. The HSE result for *m*=1.5 (open symbol) is also shown for comparison. Here, the relative displacements of both *A*O and *B*O$_2$ layers (in the two *AB*O$_3$ peroveskites) are shown together to reduce the numbers of curves. (b) The local displacements (between anions and cations) with magnetism

and (c) the local magnetic moments of Ti ions for $(LAO)_{3.5}/(PTO)_{6.5}$ with *n-n* type interfaces. The vertical olive dash lines and cyan dot lines denote the interfaces and domain walls, respectively. For reference, in (b), the red and blue dot lines denote the Ti-O and Pb-O relative displacements of PTO bulk on LAO substrate, respectively. (d) The total magnetic moment of *n-n* type SLs in both the PE/FE initial conditions as a function of the periodic thickness.

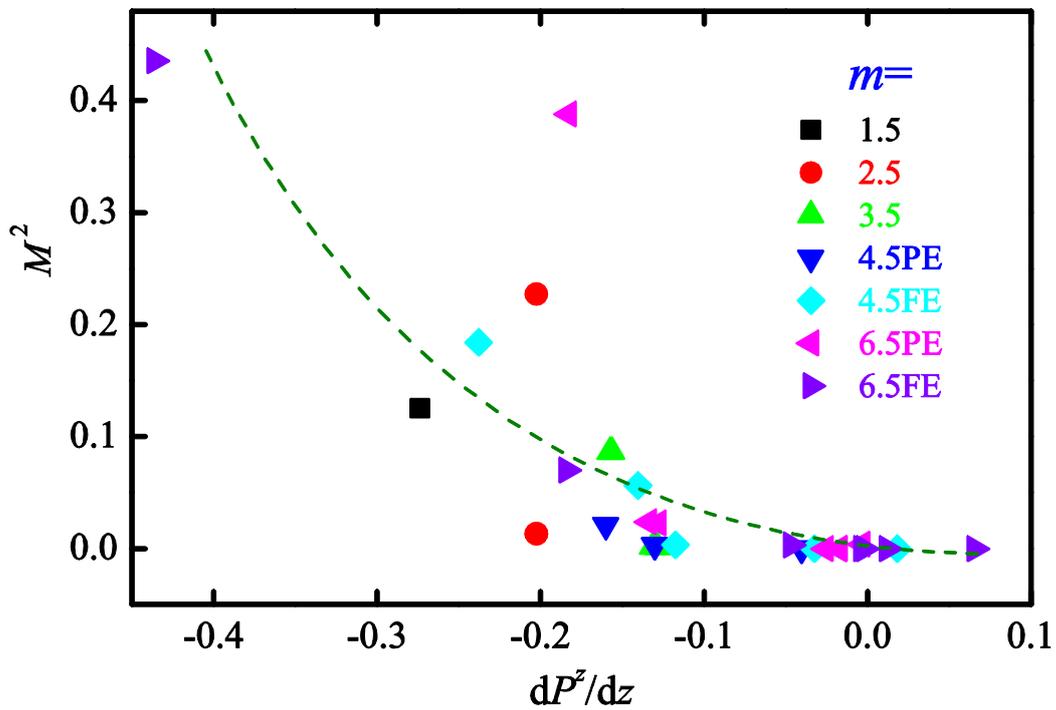

**Fig.4.** Distribution of $(dP_{iz}/dz, M_i^2)$ points in all above pure GGA calculations. For each case of a given SL, $P_{iz}$ of Ti is first fitted by a polynomial function of $z$. Here, $i$ is the unit layer number. Then the value of $dP_{iz}/dz$ is derived analytically. This process can reduce the inaccuracy due to a few discrete points as much as possible. The broken curve is just for a guide of eyes.